\documentclass{aa}
\usepackage{graphicx}
\usepackage{txfonts}
\usepackage{physics}
\usepackage[utf8]{inputenc}
\usepackage[english]{babel}
\usepackage[T1]{fontenc}
\usepackage{lmodern}
\usepackage{microtype}
\usepackage{booktabs}
\usepackage{bm}
\usepackage{graphicx}
\usepackage{amsfonts}
\usepackage{amsmath}
\usepackage{amssymb}
\usepackage{listings}
\usepackage[hidelinks]{hyperref}
\usepackage{subfigure}
\usepackage{fancyhdr}
\usepackage{mathtools}
\usepackage{natbib}
\usepackage{comment}
\usepackage{color}
\usepackage{orcidlink}
\usepackage{makecell}

\bibpunct{(}{)}{;}{a}{}{,} 

\newcommand{\msun}{\mathrm{M}_\odot}
\newcommand{\plpeak}{\textsc{PowerLaw+Peak}}
\newcommand{\betamodel}{\textsc{Beta}}
\newcommand{\evolving}{\textsc{Evolving}}
\newcommand{\plq}{\textsc{PowerLawMR}}

\begin{document} 

\titlerunning{Features in the BH mass spectrum from non-parametric analyses}
\authorrunning{S.~Rinaldi et al.}
\title{Exploration of features in the black hole mass spectrum inspired by non-parametric analyses of gravitational wave observations}

\author{
        Stefano~{Rinaldi}\inst{1,2}\orcidlink{0000-0001-5799-4155}\thanks{E-mail: stefano.rinaldi@uni-heidelberg.de}
        \and	
        Yajie~{Liang}\inst{1,3}\orcidlink{0009-0003-6181-4702}\thanks{E-mail: yajie.liang@lmu.de}
        \and
        Gabriele~{Demasi}\inst{4,5}\orcidlink{0009-0009-5320-502X}
        \and
		Michela~{Mapelli}\inst{1,2,6,7}\orcidlink{0000-0001-8799-2548}
        \and
        Walter~{Del~Pozzo}\inst{8,9}\orcidlink{0000-0003-3978-2030}
		}

\institute{
    Universit\"at Heidelberg, Zentrum f\"ur Astronomie (ZAH), Institut f\"ur Theoretische Astrophysik, Albert-Ueberle-Str. 2, 69120, Heidelberg, Germany
    \and
    Dipartimento di Fisica e Astronomia ``G. Galilei'', Università di Padova, Via Marzolo 8, 35122 Padova, Italy
    \and
    Chair of Physics Education Research, Faculty of Physics, Ludwig-Maximilians-Universität München, Munich, Germany
    \and
    Dipartimento di Fisica e Astronomia, Università degli Studi di Firenze, Via Sansone 1, Sesto Fiorentino, 50019, Italy
    \and
    INFN, Sezione di Firenze, Via Sansone 1, Sesto Fiorentino, 50019, Italy
    \and
    Universit\"at Heidelberg, Interdisziplin\"ares Zentrum f\"ur Wissenschaftliches Rechnen, Heidelberg, Germany
    \and
    INFN, Sezione di Padova, Via Marzolo 8, I--35131 Padova, Italy
    \and
    Dipartimento di Fisica ``E. Fermi'', Università di Pisa, Largo Bruno Pontecorvo 3, Pisa, I-56127, Italy
    \and
    INFN, Sezione di Pisa, Largo Bruno Pontecorvo 3, Pisa, I-56127, Italy
    }

\date{Received \today; accepted XXX}
\abstract
{Current gravitational-wave data reveal structures in the mass function of binary compact objects. Properly modelling and deciphering such structures is  the ultimate goal of gravitational-wave population analysis: in this context, non-parametric models are a powerful tool to infer the distribution of black holes from gravitational waves without committing to any specific functional form.}
{Here, we aim to quantitatively corroborate the findings of non-parametric methods with parametrised models incorporating the features found in such analyses.}
{We propose two modifications of the currently favoured \plpeak~model, inspired by non-parametric studies, and  use them to analyse the third Gravitational Wave Transient Catalogue.}
{Our analysis marginally supports the existence of two distinct, differently redshift-evolving subpopulations in the black hole primary mass function, and suggests that, to date, we are still unable to robustly assess the shape of the mass ratio distribution for symmetric ($q>0.7$) binaries.}
{}


\maketitle
\section{Introduction}
The detection of GW150914 \citep{gw150914:2015}, the first gravitational wave (GW) signal ever observed, completely changed our perspective on astrophysical compact objects \citep{astrodistGWTC1:2019}. 
The first two black holes (BHs)
whose merger was observed by the two LIGO detectors \citep{ligodetector:2015} have a mass of $36^{+5}_{-4}$ and $29^{+4}_{-4}\ \msun$ \citep{GW150914_properties:2016}. Such large masses were unexpected when compared to the BH mass distribution inferred from X-ray  binaries in the local Universe \citep{orosz:2003,oezel:2010,farr:2011}, and only a few models existed able to explain their formation from metal-poor progenitors \citep{heger:2002,mapelli:2009,mapelli:2010,belczynski:2010,fryer:2012,mapelli:2013,spera:2015}. 

During the three observing runs concluded to date, the LIGO-Virgo-KAGRA (LVK) collaboration observed almost one hundred GW events making use of the four interferometers currently active -- the two LIGO instruments, Virgo \citep{virgodetector:2015} and KAGRA \citep{kagradetector:2013} -- and this number is set to grow in the next future, with the new GW signals that are currently being detected during the currently ongoing fourth observing run (O4).

From an astrophysical perspective, 
characterising the BH population is of the utmost importance: the evolution of a star leaves an imprint on the properties of its compact remnant. Hence, by reconstructing the properties of the BH population, we can constrain the life and death of their progenitor stars, and we can explore the many proposed formation channels of binary compact objects: isolated binary and triple evolution \citep[e.g.,][]{bethe:1998,eldridge:2016,kruckow:2018,iorio:2023,fragos:2023,briel:2023}, or dynamical formation in star clusters \citep[e.g.,][]{portegieszwart:1999,rodriguez:2016,mapelli:2016,askar:2017,samsing:2018,dicarlo:2020,fragione:2020,arcasedda:2021,dallamico:2021,romero:2022,torniamenti:2024,Mahapatra:2025}
and AGN disks \citep[e.g.,][]{mckernan:2012,mckernan:2018,tagawa:2020,tagawa:2021,samsing:2022,vaccaro:2024}.

The staggering complexity of astrophysical models, along with the fact that their computation is still relatively slow compared to the number of evaluations required by Monte Carlo methods, makes them ill-suited for a fine-grained inference of the relevant parameters via stochastic sampling. Albeit some techniques to circumvent this issue (including but not limited to machine learning interpolants) are currently being explored 
\citep{colloms:2025, Karathanasis:2023, Afroz:2024}, most of the available literature \citep[e.g.,][]{fishbach:2017,talbot:2018,astrodistGWTC1:2019,astrodistGWTC2:2021,astrodistGWTC3:2023,farah:2023,gennari:2025,li:2025} relies on parametrised models inspired by the expected underlying physics. In particular, the model that is currently favoured by the publicly available observations is the \textsc{PowerLaw+Peak} model, described in Appendix~B1.b of \citet{astrodistGWTC3:2023}. This model parametrises the mass spectrum as a power-law distribution inspired by the stellar initial mass function \citep{salpeter:1955,kroupa:2001} with a low- and high-mass cut-off \citep{fishbach:2017,wysocki:2019} superimposed to a Gaussian feature at around 35-40 $\msun$ \citep{talbot:2018}. In this model, the mass ratio distribution is parametrised as a power-law, and the redshift distribution is assumed to be uniform in co-moving volume with a merger rate density evolution following \citet{madau:2014}:
\begin{equation}
    p(z) \propto \frac{\dd V_c}{\dd z} (1+z)^{\kappa-1}\,.
\end{equation}

Another approach, alternative to parametrised models, makes use of the so-called `non-parametric models', data-driven models requiring minimal mathematical assumptions and described by a countable but potentially infinite number of parameters whose functional form is flexible enough to approximate arbitrary probability densities making use of minimal mathematical assumptions. Some examples in this sense are models making use of autoregressive processes \citep{callister:2024}, Dirichlet process Gaussian mixture models \citep{rinaldi:2022:hdpgmm}, finite Gaussian mixtures \citep{tiwari:2021}, Gaussian processes \citep{li:2021}, reversible jump Markov chain Monte Carlo \citep{toubiana:2023}, cubic splines \citep{edelman:2022}, and binned approaches \citep{ray:2023,heinzel:2024}. The strength of this different approach is that all the available information comes from the data alone and does not require any specific assumption regarding the shape of the underlying distribution: doing so, features in the BH mass spectrum will arise naturally, without the need to include them by hand in the model itself. The downside is that, being the non-parametric models completely agnostic and uninformed on the physics governing the phenomena behind the inferred distribution, a direct physical interpretation of the inferred non-parametric distribution is impossible: at the current stage, a non-parametric reconstruction of the BH mass function can only be used to guide the further development of parametrised and/or astrophysical models.

Here we investigate some features of the BH distribution that are found in several non-parametric studies, introducing variations in the three-dimensional population model -- namely, changing the mass ratio distribution and the mass dependency of the redshift distribution -- and we assess their robustness with respect to the `vanilla' \plpeak, used as benchmark.
The paper is structured as follows: in Section~\ref{sec:motivations} we review some non-parametric results, presenting the features we want to investigate. We present our methodology in Section~\ref{sec:methods}. Section~\ref{sec:results}  describes our findings; specifically, Section~\ref{sec:plmassratio} explores the mass ratio distribution, whereas Section~\ref{sec:plredshift} reports the possible presence of a redshift-dependent primary mass distribution. Section~\ref{sec:conclusions} we outline a brief discussion of our findings and summarise our conclusions.

\section{Motivation}\label{sec:motivations}
Several works investigate the BH mass function using non-parametric methods: the brief overview presented in this Section is not meant to be exhaustive, but rather intends to focus on the two specific features that we want to discuss in this paper. 

From an astrophysical perspective, the mass ratio distribution of BBHs is heavily affected by a variety of processes: among others, we mention the mass transfer between the two components of a binary system and the pairing mechanisms of compact objects in dense environments such as clusters.
A robust characterisation of such distribution of BBHs is therefore paramount to understanding the physics behind compact binary formation.
The \plpeak~model describes the mass ratio distribution using a power-law
\begin{equation}\label{eq:q_pl}
    p(q)\propto q^{\beta}\,,
\end{equation}
whose index $\beta$ is likely positive: $\beta = 1.1^{+1.7}_{-1.3}$ \citep{astrodistGWTC3:2023}. \footnote{Throughout the text, the quoted uncertainties correspond to the 90\% credible interval.}
With such index, mainly driven by the lack of asymmetric binaries among the available GW events, the mass ratio distribution will inherently favour more symmetric objects.
In the case of flexible non-parametric models, the mass ratio distribution is not required to have a maximum for symmetric binaries: multiple non-parametric studies hint at the fact that the mass ratio distribution does not have its maximum for $q=1$ \citep{edelman:2023,rinaldi:2024:m1qz, Sadiq:2024,heinzel:2024b}. 
The second point we want to highlight is the evolution of the primary mass distribution with redshift. Several works investigate this topic \citep[e.g.,][]{fishbach:2021,vanson:2022,astrodistGWTC3:2023,gennari:2025,lalleman:2025} using parametric or weakly parametrised models, whereas other works explore the correlation between these two parameters using a non-parametric method \citep{ray:2023,rinaldi:2024:m1qz,heinzel:2024, Sadiq:2025}. 
So far, most of the investigations with (weakly) parametrised models has returned inconclusive null results \citep[e.g.,][]{ray:2023,gennari:2025} favouring stationary models over their evolving counterparts.
A proper accounting of selection effects is paramount in population studies to avoid biases \citep{essick:2024}: in particular, non-parametric methods are not suitable for extrapolation outside the region where no data are available due to their agnosticity. Therefore, the lack of low-mass, high redshift systems highlighted by non-parametric analyses is most likely to be ascribed to the limited sensitivity of our detectors.

When we consider the opposite scenario from a detectability point of view, low-redshift, high-mass systems are expected to produce the loudest, and thus easier to detect, signals.
Nonetheless, these objects are only observed from redshift $z\sim 0.2-0.3$, whereas low-mass BHs are detected also at smaller redshifts. 
While investigating the primary mass-redshift plane, \citet[][see their Figure~1]{heinzel:2024b}, \citet[][see their Figure~4]{rinaldi:2024:m1qz}, 
and \citet[][see their Figure~6]{lalleman:2025} find results that, albeit not strongly supporting a change in redshift evolution, mildly suggest a different behaviour for the 10 $\msun$ feature and the 35 $\msun$ feature at redshifts $z < 0.25$, before selection effects kicks in. Qualitatively, the two features appear to grow at a different rate with redshift: for example, in Figure~6 of \citet{lalleman:2025}, the authors find that the large uncertainties on the power-law index $\alpha_z$ are consistent with a constant slope, but the median value shows a small kink at around $35\ \msun$.
A simple explanation for this different behaviour in the two mass regimes might the difference in relative abundance of the two features, with the lack of low-redshift, high-mass black holes being due to Poissonian fluctuations.
In this work, however, we explore the hypothesis where the different redshift trend suggests that we are in presence of two different BH populations, potentially coming from two different formation channels. Such possibility has been investigated by \citet{vanson:2022} making use of the GW observations up to the first half of the third observing run \citep{GWTC2:2021}, finding marginal evidence in favour of this hypothesis.

In what follow, we will make use of the BBHs observed in the first three LVK observing runs to test these two hypotheses suggested by non-parametric studies.

\section{Summary of statistical methods}\label{sec:methods}
Before describing the modifications to the \plpeak~models introduced in this work, we briefly summarise the statistical method used to infer the parameters of the astrophysical BH distribution. This derivation was first presented by \citet{mandel:2018}, expanding the concept originally introduced by \citet{loredo:2004}. Our objective is to characterise the posterior probability distribution for a set of parameters describing the astrophysical distribution of BHs, $p(\theta|\Lambda)$. Here $\theta$ denotes the parameters describing the BBH, such as $\mathrm{M}_1$, $q$ and $z$, whereas $\Lambda$ can include, for example, the spectral index of the primary mass power-law distribution and the mean of the Gaussian peak. The observed data is denoted with $\mathbf{d} = \{d_1,\ldots,d_n\}$ (the index labels the different GW events) and the fact that only a fraction of all the existing BBH coalescences is detectable is indicated with $\mathbb{D}$. The likelihood reads
\begin{equation}
p(\mathbf{d}|\Lambda, \mathbb{D}) = \prod_i \int p(d_i|\theta_i,\mathbb{D})p(\theta_i|\Lambda,\mathbb{D})\dd \theta_i\,,
\end{equation}
under the assumption that the events are independent and identically distributed and that the individual event likelihood is independent of the astrophysical parameters $\Lambda$. For each term of the product, the integrand can be rewritten as 
\begin{equation}
\frac{p(\mathbb{D}|d_i,\theta_i)p(d_i|\theta_i)}{p(\mathbb{D}|\theta_i)}\frac{p(\mathbb{D}|\theta_i)p(\theta_i|\Lambda)}{p(\mathbb{D}|\Lambda)} = \frac{p(d_i|\theta_i)p(\theta_i|\Lambda)}{p(\mathbb{D}|\Lambda)}\,,
\end{equation}
where we simplified $p(\mathbb{D}|\theta_i)$ between the two terms and assumed that, by definition, the detectability of the observed events is equal to 1.
The term $p(\mathbb{D}|\Lambda)$ is the detectability fraction of the population, often denoted with $\alpha(\Lambda)$:
\begin{equation}\label{eq:detectability_def}
p(\mathbb{D}|\Lambda) = \int p(\mathbb{D}|\theta)p(\theta|\Lambda)\dd \theta \equiv \alpha(\Lambda)\,.
\end{equation}
In conclusion, the likelihood reads
\begin{equation}\label{eq:full_likelihood}
    p(\mathbf{d}|\Lambda, \mathbb{D}) = \prod_i \frac{p(d_i)}{\alpha(\Lambda)}\int \frac{p(\theta_i|d_i)p(\theta_i|\Lambda)}{p(\theta_i)}\dd\theta_i\,.
\end{equation}

This derivation, closely following \citet{mandel:2018}, can be extended to the case in which the considered astrophysical model $p(\theta|\Lambda)$ is a weighted superposition of models:
\begin{equation}
p(\theta|\Lambda) = \sum_j w_j p_j(\theta|\Lambda_j)\,.
\end{equation}
Here $p_j(\theta|\Lambda_j)$ denotes one specific component of the mixture model (i.e., the Gaussian component of the \plpeak~model) and $\Lambda_j$ the relevant subspace of the global parameter space $\Lambda$ (i.e, the mean and variance of the Gaussian). With this in mind, the integral in Eq.~\eqref{eq:full_likelihood} can be rewritten as
\begin{equation}
\int \frac{p(\theta_i|d_i)p(\theta_i|\Lambda)}{p(\theta_i)}\dd\theta_i = \sum_j w_j \int \frac{p(\theta_i|d_i)p_j(\theta_i|\Lambda_j)}{p(\theta_i)}\dd\theta_i
\end{equation}
as well as $\alpha(\Lambda)$:
\begin{multline}
\alpha(\Lambda) = \int p(\mathbb{D}|\theta)p(\theta|\Lambda) \dd\theta = \sum_j w_j \int p(\mathbb{D}|\theta)p_j(\theta|\Lambda_j)\dd \theta \\ \equiv \sum_jw_j \alpha_j(\Lambda_j)\,.
\end{multline}
Therefore, Eq.~\eqref{eq:full_likelihood} in the case of a mixture model can also be expressed as
\begin{multline}
p(\mathbf{d}|\Lambda) =\\ \prod_i \sum_j \frac{w_j \alpha_j(\Lambda_j)}{\sum_k w_k\alpha_k(\Lambda_k)}\frac{p(d_i)}{\alpha_j(\Lambda_j)}\int \frac{p(\theta_i|d_i)p_j(\theta_i|\Lambda_j)}{p(\theta_i)}\dd\theta_i\\ = \prod_i \sum_j \frac{w_j \alpha_j(\Lambda_j)}{\sum_k w_k\alpha_k(\Lambda_k)} p(d_i|\Lambda_j)\,,
\end{multline}
a weighted superposition of the likelihoods for each mixture components where the relative weights accounts also for the selection effects.

In this work, we make use of the publicly available GW events released by the LVK collaboration in GWTC-2.1 \citep{GWTC21:2024} and GWTC-3 \citep{GWTC3:2021} obtained via GWOSC\footnote{\url{https://gwosc.org}}. In particular, we use the second, most recent version of the parameter estimation samples (v2).
We selected the events with the same criteria used in \citet{astrodistGWTC3:2023} -- false alarm rate (FAR) < 1 yr$^{-1}$ -- for a grand total of 69 BBHs.
The selection effects are accounted for using the search sensitivity estimates for O1, O2 and O3 \citep[][v2]{sensitivityestimate:2023} released along with GWTC-3.

Concerning the BH parameters that will be included in our analysis, we decided to work with a subset of the 9 astrophysically relevant parameters (two mass parameters, 6 spin parameters and one distance parameter), considering the primary mass, the mass ratio and the redshift. The spin parameters can be marginalised out from the likelihood $p(\mathbf{d}|\Lambda)$ under the following assumption:
\begin{itemize}
\item The astrophysical model $p(\theta|\Lambda)$ is separable as $p(\mathrm{M}_1,q,z|\Lambda)p(\mathbf{s}|\Lambda)$. This is possible if we assume a Cartesian or polar parametrisation for the spin parameters instead of the $(\chi_\mathrm{eff},\chi_\mathrm{p})$ and angles parametrisation used, for example, in \cite{astrodistGWTC3:2023};
\item The individual event posterior distribution $p(\theta_i|d_i)$ does not show correlations among $\mathrm{M}_1$, $q$, and $z$ and the spin parameters;
\item The spin distribution $p(\mathbf{s}|\Lambda)$ is the same as the one used to generate the mock signals used in the LVK search sensitivity estimates \citep{sensitivityestimate:2023}. Doing this, the Monte Carlo sum used to estimate the detectability integral in Eq.~\eqref{eq:detectability_def} is unbiased \citep{vitale:2022}.
\end{itemize}
These hypotheses are satisfied -- with a degree of approximation for the second\footnote{The largest Pearson correlation coefficient averaged on the events is between the mass and the spin magnitude of the primary object, $\langle r_{M_1s_1}\rangle = 0.12$} -- therefore we will consider, in our analysis, $\mathrm{M}_1$, $q$, and $z$.

\section{Results}\label{sec:results}
We now present the modifications to the \plpeak~model we introduce and the results of the analyses made with such models. The mass ratio distribution inference is performed using \textsc{GWPopulation}\footnote{Publicly available at \url{https://github.com/ColmTalbot/gwpopulation} and via \texttt{pip}.} \citep{talbot:2019}, whereas the primary mass distribution inference is build upon \textsc{RayNest}\footnote{Publicly available at \url{https://github.com/wdpozzo/raynest} and via \texttt{pip}.}.
\subsection{Mass ratio distribution}\label{sec:plmassratio}
To explore different possible behaviours of the mass ratio distribution, we replace the power-law model (Eq.~\ref{eq:q_pl}) with a Beta distribution with one additional degree of freedom, characterized by $\alpha_q$ and $\beta_q$:
\begin{equation}
        p(q) \propto q^{\alpha_q -1} (1-q)^{\beta_q - 1}\,.
\end{equation}
The power-law is a special case of the Beta distribution with $\beta_q = 1$ and $\alpha_q -1 \equiv \beta$ from Eq.~\eqref{eq:q_pl}. We assume a uniform prior for both of these parameters: $\mathcal{U}(0, 30)$.
In what follows we will denote this model with \betamodel~and the standard power-law one with \plq.

\begin{figure*}
    \centering
    \subfigure[]{
    \includegraphics[width=0.75\columnwidth]{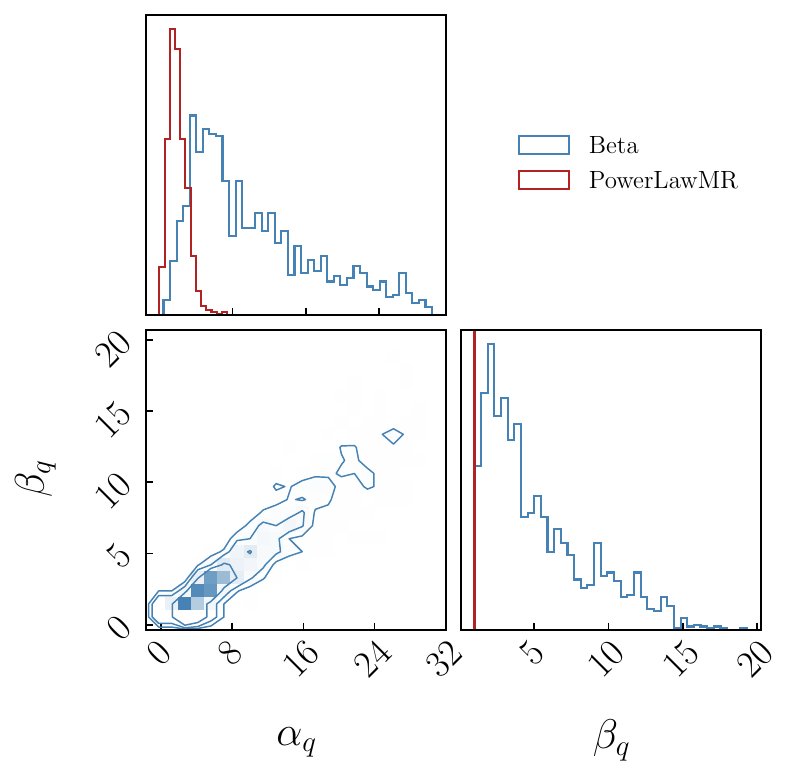}\label{fig:q_paras}
    }
    \subfigure[]{
    \includegraphics[width=0.9\columnwidth]{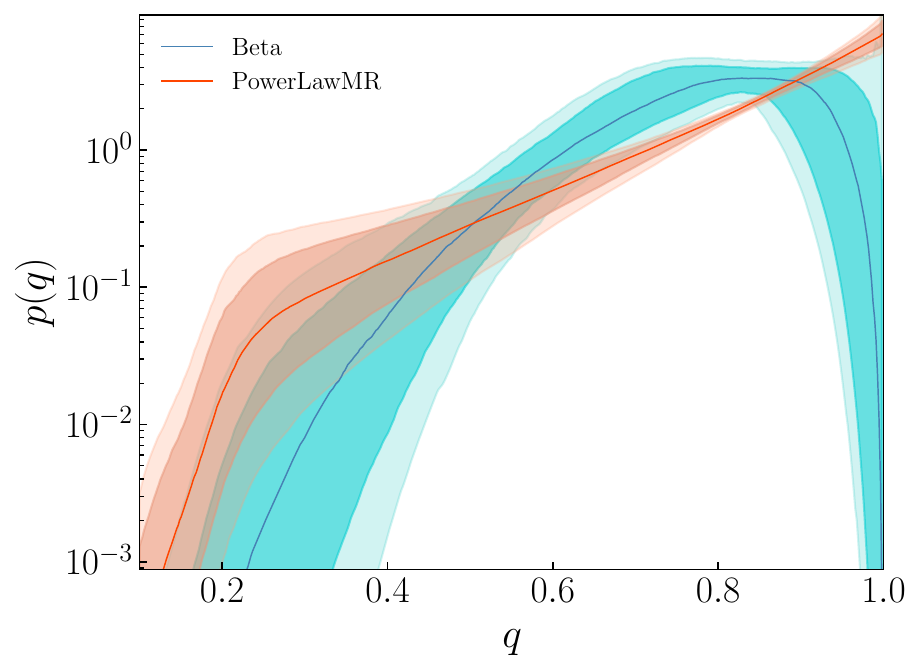}\label{fig:q_m}
    }
    \caption{Left: Posterior distribution for the $\alpha_q$ and $\beta_q$ parameters of the \betamodel~model (blue) compared with the corresponding $\beta + 1$ parameter of the \plq~model. Right: Marginal posterior probability density for the mass ratio $q$ using the two different models described in Sec.~\ref{sec:plmassratio}. The shaded areas correspond to the 68\% and 90\% credible regions.}
\end{figure*}

Figure~\ref{fig:q_paras} shows the marginal posterior distributions and correlation of $\alpha_q$ and $\beta_q$.
In particular, the inferred value of $\beta_q = 4.1^{+8.3}_{-2.7}$ suggests a deviation from the simple power-law behaviour that would be denoted by a posterior on $\beta_q$ consistent with 1.
As shown in Fig.~\ref{fig:q_m}, the inferred mass ratio distribution most likely has a plateau between $q\sim 0.75$ and $q\sim 0.9$, with a cut-off for almost-symmetric binaries -- this, however, without excluding the possibility for the mass ratio distribution to have its maximum at $q = 1$. The log Bayes' factor among \plq~and \betamodel~is $\ln\mathcal{B} = 0.94\pm 0.5$, marginally favouring the first model. Now, since the Bayes' factor contains a penalty factor -- sometimes referred to as the ``Occam's razor'' -- disfavouring a priori the models with more parameters through a larger prior volume, we argue that such marginal evidence can be due to the additional parameter $\beta_q$ present in the \betamodel~model. Therefore, we reduced the dimensionality of the parameter space for \betamodel~fixing $\beta_q$ to its maximum a posteriori value, $\beta_q = 4.1$, and repeating the inference. In this case, we get $\ln\mathcal{B} = 0.22 \pm 0.5$ in favour of \plq: this value is consistent, within the associated uncertainty, with equally favoured models. The different log Bayes' factors we computed are summarised in Table~\ref{tab:bayesfactors}.

\begin{table}[]
    \centering
    \caption{Log Bayes' factors among the different models considered in Sec.~\ref{sec:plmassratio}.}
    \begin{tabular}{cc}
    \toprule
       Models  & $\ln{\mathcal{B}^{A}_{B}}$ \\
       \midrule
       \makecell{(A) \plq; \, (B) \betamodel}  &  $0.94 \pm 0.5$\\
       \midrule
       \makecell{(A) \betamodel, $\beta_q = 4.1$; \, (B) \betamodel} &  $0.72 \pm 0.5$\\
       \midrule
       \makecell{(A) \plq; \, (B) \betamodel, $\beta_q = 4.1$} &  $0.22 \pm 0.5$\\
    \bottomrule
    \end{tabular}
    \label{tab:bayesfactors}
\end{table}

The reason why two models with very distinct behaviours in the symmetric binary regime ($q\gtrsim0.7$) similarly explain the available data should be looked for in the amount of information carried by the posterior distribution with respect to the prior and, most importantly, how this is distributed as a function of the variable of interest, the mass ratio $q$. We will denote the prior distribution with $\Pi(q)$, which derives from the bounded uniform prior on both $\mathrm{M}_1$ and $\mathrm{M}_2$,
\begin{equation}
    \Pi(q) \propto \int_{\mathrm{M}_\mathrm{min}}^{\mathrm{M}_\mathrm{max}} \mathrm{M}_1 \Theta\qty(q\mathrm{M}_1 - \mathrm{M}_\mathrm{min}) \dd \mathrm{M}_1\,.
\end{equation}
Here $\Theta(\cdot)$ is the Heaviside step function, whereas $\mathrm{M}_\mathrm{min}$ and $\mathrm{M}_\mathrm{max}$ denote the minimum and maximum possible BH mass both for the primary and secondary object of a binary system. The posterior distribution, denoted with $\mathcal{P}(q)$, is obtained fitting a Dirichlet process Gaussian mixture model to the mass ratio posterior samples for each GW event in GWTC-3.

\begin{figure*}
    \centering
    \subfigure[Mass ratio]{
    \includegraphics[width=0.9\columnwidth]{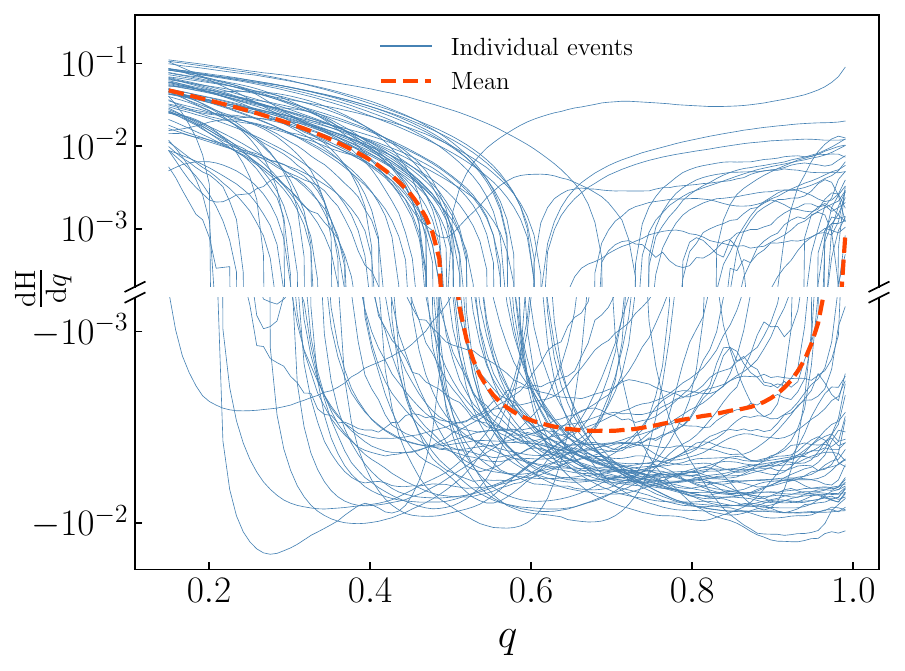}\label{fig:information_q}
    }
    \subfigure[Primary mass]{
    \includegraphics[width=0.9\columnwidth]{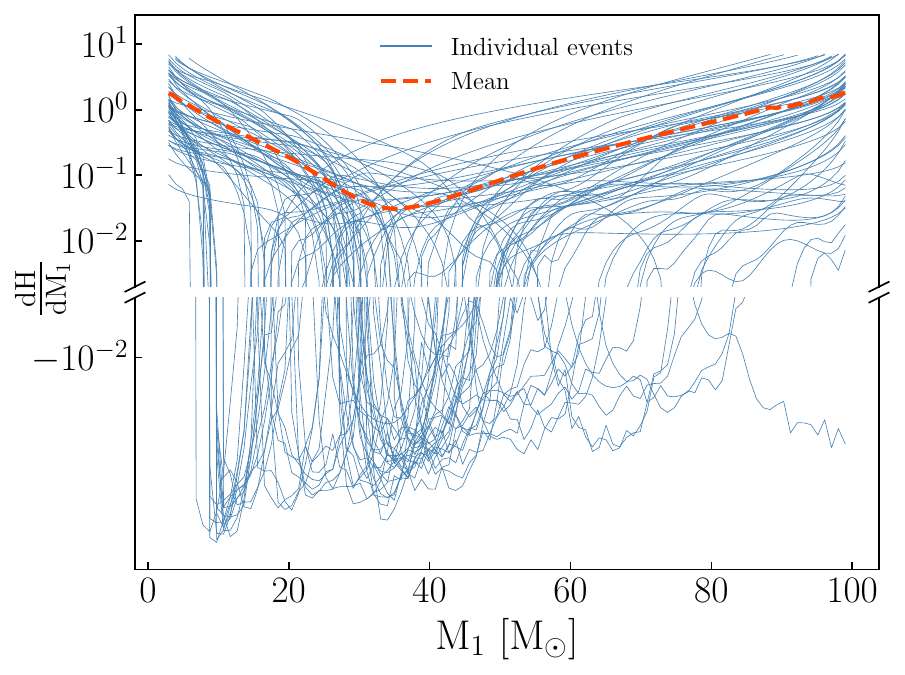}\label{fig:information_m}
    }
    \caption{Differential information entropy between prior and marginal posterior as a function of mass ratio (left) and primary mass (right). Each thin blue solid line represents one individual GW event, whereas the thick dashed red line marks the average differential information entropy. Note that the top and bottom part of both panels have different limits.}
\end{figure*}

Using these functions and following the definition of information entropy given by \citet{shannon:1948}, we define the differential information $\dd \mathrm{H} /\dd q$ as
\begin{equation}
    \frac{\dd \mathrm{H}}{\dd q} = \Pi(q)\ln(\frac{\Pi(q)}{\mathcal{P}(q)})\,.
\end{equation}
This function characterises the information content of the posterior with respect to the prior, diverging in all these regions of the parameter space where $\Pi(q)$ is finite and $\mathcal{P}(q) \to 0$ and approaching 0 where $\Pi(q)\simeq \mathcal{P}(q)$ (hence where the likelihood is non-informative). In this second case, the only available source of information for the population inference is the functional form of the astrophysical distribution, thus marking a hyperprior-driven\footnote{The prior driving the inference is not the one used to perform the parameter estimation of individual binaries, but rather the functional form a priori assumed to describe the inferred distribution.} area of the parameter space.

Figure~\ref{fig:information_q} shows the differential information entropy as a function of the mass ratio $q$. For the vast majority of the available GW events, most of the information is accumulated for $q <0.6$, with comparably less information carried for symmetric binaries (GW190412 being a notable exception). This is in agreement with our finding that the two models, \plq~and \betamodel, describe differently the astrophysical population of mostly symmetric BHs despite yielding the same evidence. In practice, the two models are mainly calibrated in the $q<0.6$ region, and then extrapolated towards symmetric binaries. For comparison, we also report the same plot for $\mathrm{M}_1$: in this case, the posterior distributions are more informative -- in the sense that they individually exclude a larger fraction of the prior volume -- and overall the average differential information entropy is almost two orders of magnitude larger, not showing any region with a clear lack of information.

In light of the findings discussed in this Section, we believe that the GW events currently publicly available are not sufficiently informative about symmetric binaries, and therefore care must be taken while drawing conclusions on the shape of the mass ratio distribution based on these events.

\subsection{Redshift evolution of subpopulations in the mass spectrum}\label{sec:plredshift}
We now turn our attention to the evolution of the black hole mass function with redshift. The model used for the analysis presented in this section follows the one proposed by  \citet{vanson:2022}, where the authors explore the possibility of a separate redshift evolution for the two components (power-law and peak) of the mass spectrum. We will therefore write the probability distribution as

\begin{multline}
    p(\mathrm{M}_1,q,z) = (1-\lambda)\,p(q|\mathrm{M}_1)\,\mathrm{PL}(\mathrm{M}_1)\,p(z|\kappa_\mathrm{PL})\\ + \lambda \,p(q|\mathrm{M}_1)\,\mathcal{N}(\mathrm{M}_1)\,p(z|\kappa_\mathrm{Peak})\,.
\end{multline}
Here, $\mathrm{PL}(\mathrm{M}_1)$ denotes the tapered power-law from the \plpeak~model (which we will use as benchmark) and $\mathcal{N}(\mathrm{M}_1)$ denotes the Gaussian distribution.
For both components, the mass ratio distribution follows the power-law in Eq.~\eqref{eq:q_pl} used in \citet{astrodistGWTC3:2023} with a common exponent $\beta$ (i.e., the two components share an identical mass ratio distribution). The redshift distribution can be expressed as
\begin{equation}
p(z|\kappa) \propto \mathcal{R}(z|\kappa) \,\frac{\dd V_c}{\dd z} \,(1+z)^{-1} \propto \frac{\dd V_c}{\dd z} \,(1+z)^{\kappa-1}\,,
\end{equation}
where the rate is parametrised as
\begin{equation}
    \mathcal{R}(z|\kappa) = \mathcal{R}_0\,(1+z)^\kappa\,,
\end{equation}
with an independent spectral index for each component ($\kappa_\mathrm{PL}$ and $\kappa_\mathrm{Peak}$). We will denote this model as \evolving.
In what follows, we will make use of the same parametrisation and notation used in \citet{astrodistGWTC3:2023}, as well as the priors on the model parameters. We will assume the same prior for both $\kappa_\mathrm{PL}$ and $\kappa_\mathrm{Peak}$, $\mathcal{U}(-4,8)$.
To ensure the robustness of our results against possible biases due to the exclusion of the spin parameters we run the analysis with both the \evolving~and \plpeak~models ourselves instead of comparing our findings directly with the official LVK data release.

\begin{figure}
    \centering
    \includegraphics[width=0.95\columnwidth]{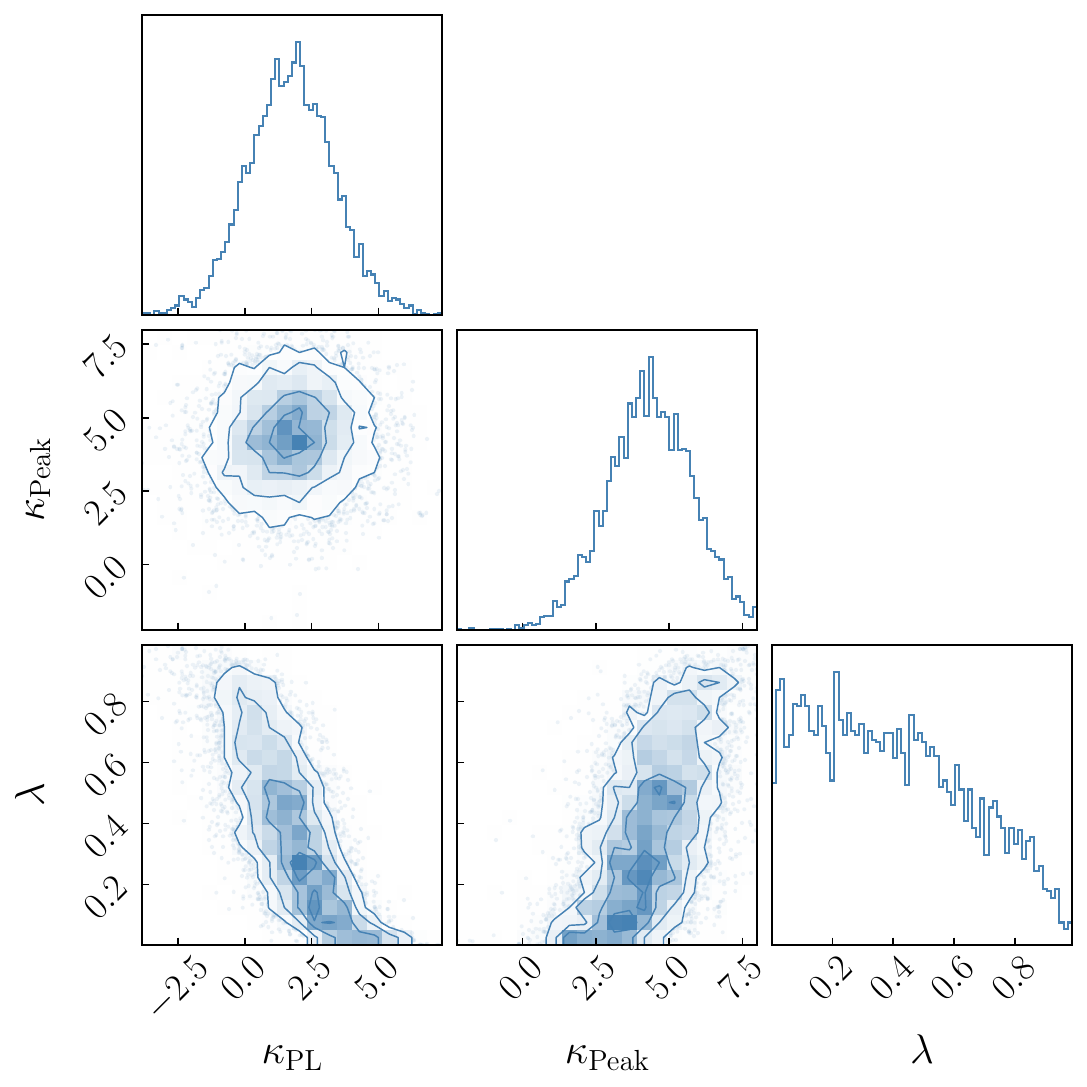}
    \caption{Joint posterior distribution for $\kappa_\mathrm{PL}$, $\kappa_\mathrm{Peak}$ and $\lambda$ for the \evolving~model.}
    \label{fig:lambda_hist}
\end{figure}

The first, most important result is that our \evolving~model yields a $\ln\mathcal{B} = 2.12 \pm 0.1$ in its favour with respect to the \plpeak~model, confirming and strengthening the marginal evidence found by \citet{vanson:2022} for this specific model of redshift evolution. 
On top of this, all parameters that are shared with the benchmark model agree with the values reported by  \citet{astrodistGWTC3:2023} with the only exception of $\lambda$, whose marginal posterior distribution is reported in the bottom-right panel of Figure~\ref{fig:lambda_hist} and which will be discussed below. Despite not ensuring the complete absence of biases in the analysis, the agreement between our posteriors and the LVK ones -- with the exception of the $\lambda$ parameter -- do not highlight any blatant inconsistency with a model that includes a non-fixed spin distribution.

\begin{figure*}
    \centering
    \subfigure[]{    \includegraphics[width=0.9\columnwidth]{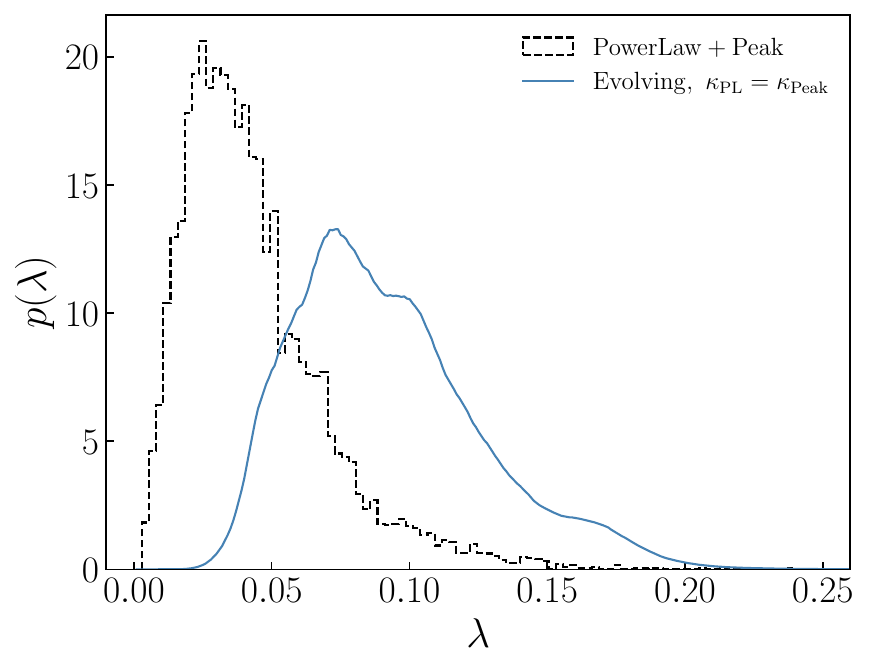}\label{fig:lambda_cond}
    }
    \subfigure[]{
    \includegraphics[width=0.9\columnwidth]{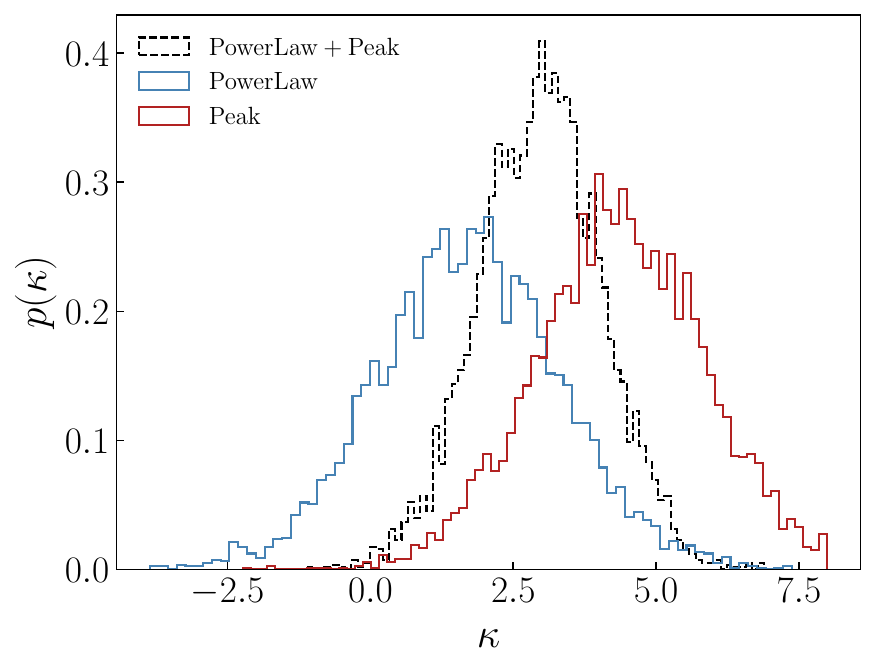}\label{fig:kappa_hist}
    }
    \caption{Left: Marginal posterior distribution for $\lambda$ for the \evolving~model conditioned on $\kappa_\mathrm{PL} = \kappa_\mathrm{Peak}$ and for the \plpeak~model. Right: Marginal posterior distribution for the redshift spectral index $\kappa$ for the PowerLaw, Peak and \plpeak~population model.}
\end{figure*}

The difference mentioned above stems from the fact that $\lambda$ is defined differently between the two models: whereas in the \plpeak~model it measures the relative weight of the two mass components only, in the \evolving~scenario the $\lambda$ parameter accounts also the different redshift evolution, thus correlating with the $(\kappa_\mathrm{PL},\kappa_\mathrm{Peak})$ parameters (see Figure~\ref{fig:lambda_hist}). Therefore, a direct comparison between the posteriors of the $\lambda$ parameter of the \plpeak~model and the one of the \evolving~model is not straightforward. Nonetheless, making use of a continuous approximant based on the Gaussian Mixture Model \citep{rinaldi:2024:figaro} for this three-dimensional posterior, we obtained the marginal posterior distribution for $\lambda$ under the condition $\kappa_\mathrm{PL} = \kappa_\mathrm{Peak}$ reported in Figure~\ref{fig:lambda_cond}. This probability density is, as expected, more in agreement with the \plpeak~result for the same parameter than the unconstrained marginal posterior.

Figure~\ref{fig:kappa_hist} shows the three marginal posterior distributions for the redshift indexes $\kappa_\mathrm{PL} = 1.7^{+1.5}_{-1.6}$, $\kappa_\mathrm{Peak} = 4.3^{+1.4}_{-1.4}$ (\evolving) and $\kappa = 3.0^{+1.0}_{-1.0}$ (\plpeak). We see that, albeit the two posterior distributions for the \evolving~model parameters are still partially consistent with each other.

To quantify their consistency, we compare the two marginal distributions $p(\kappa_\mathrm{PL}|\mathbf{d})$ and  $p(\kappa_\mathrm{Peak}|\mathbf{d})$ using the Hellinger distance \citep{hellinger:1909}, which is defined as
\begin{equation}
    H_D(p,q) = \qty(1-\int\sqrt{p(x)q(x)}dx)^{1/2}\,.
\end{equation}
For these distributions (under the assumption that $\kappa_\mathrm{PL}=\kappa_\mathrm{Peak}$) we get $H_D = 0.55$, indicating that the difference between the two distribution is significant but inconclusive. This is in agreement with the marginal evidence found in favour of two different redshift evolutions.
The results we find are consistent with the analysis presented in \citet{lalleman:2025}, where the authors use a more flexible version of the \evolving~model presented here: the marginal evidence in favour of this model reported in this Section is in agreement with the non-conclusive hint to a change in redshift index at $\sim 35\ \msun$ reported in their Figure~6.
Moreover, the posterior distribution for $\kappa$ matches exactly the overlapping region of the other two marginal distributions, qualitatively suggesting that the redshift distribution inferred with the \plpeak~model is simply the average of the evolution of the two separate populations. 

\begin{figure*}\label{fig:redshiftevol}
    \centering
    \subfigure[PDF]{
    \includegraphics[width=0.9\columnwidth]{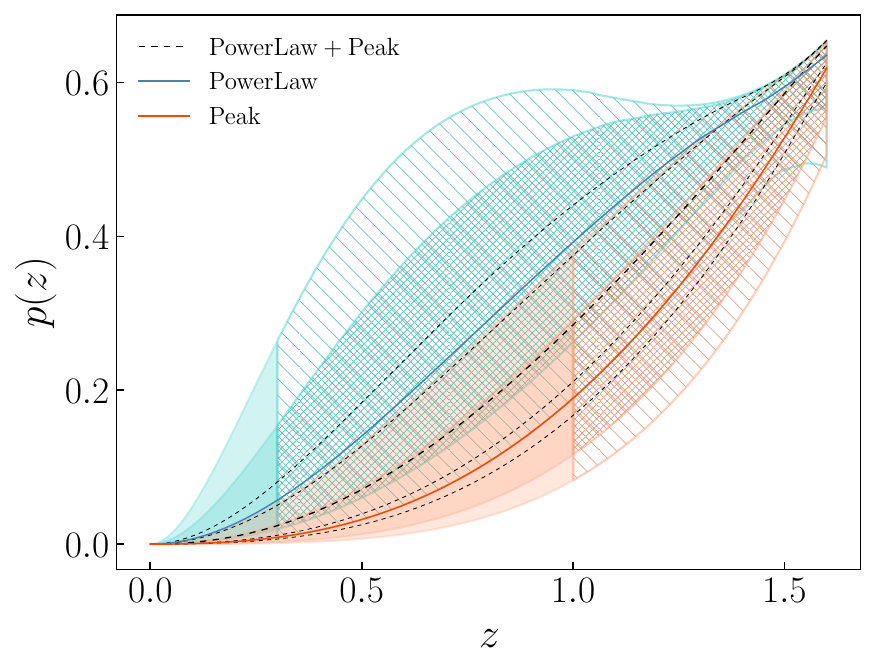}\label{fig:pdf_plot}
    }
    \subfigure[Merger rate density]{
    \includegraphics[width=0.9\columnwidth]{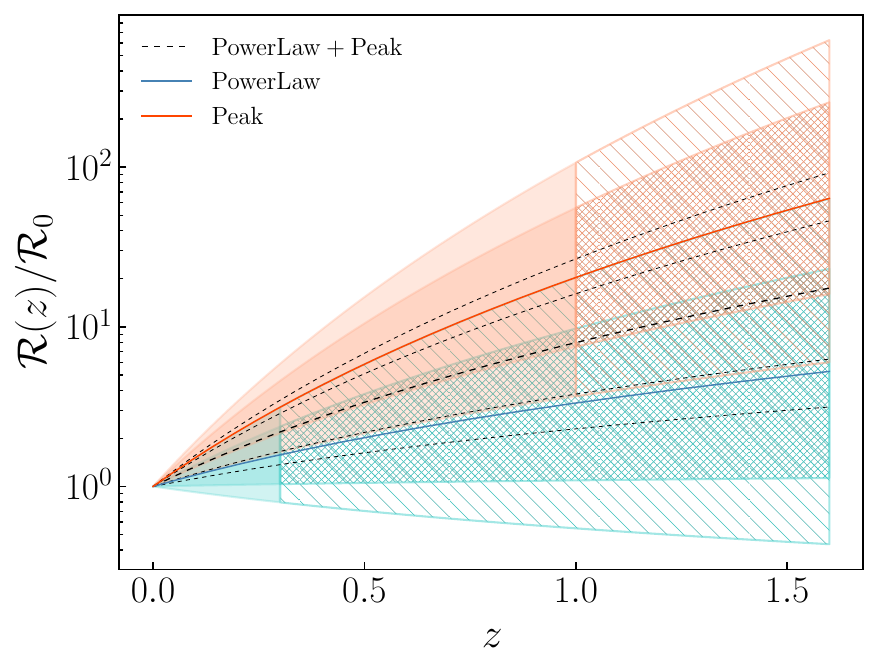}\label{fig:rate_plot}
    }
    \caption{Left: probability density for the source redshift. Right: Merger rate density as a function of redshift. In both panels, the coloured regions refer to the two components of the \evolving~model, whereas the areas delimited by the dashed black line refers to the common evolution of the \plpeak~model. The hatched areas mark the extrapolation beyond detector horizon for the corresponding mass range.}
\end{figure*}

It is important to keep in mind, however, that the region corresponding to low-mass, high-redshift BBHs is almost entirely vetoed by the detector sensitivity: therefore, any conclusion drawn on the low-mass end of the BH spectrum is driven by the objects observed in the very local Universe, $z\lesssim 0.3$ (see Fig.~\ref{fig:redshiftevol}). Future observations with higher sensitivity -- like the fourth and fifth observing runs, as well as the third generation detectors, Einstein Telescope and Cosmic Explorer -- will help assessing whether this observed different evolution between the two channels is actually a feature of the BH population.

\section{Summary}\label{sec:conclusions}
In this paper we introduced and investigated two possible modifications of the $\plpeak$~model used by  \citet{astrodistGWTC3:2023} to describe the BH mass spectrum to accommodate new features found by non-parametric studies. Our findings can be summarised in two main points:
\begin{itemize}
    \item The amount of information currently encoded in GWTC-3 regarding symmetric binaries is not sufficient to constrain the mass ratio distribution shape for $q > 0.7$. This suggests that the inferred behaviour of the mass ratio distribution for symmetric binaries is mostly driven by the few detected asymmetric systems and the specific shape of the chosen model;
    \item There is marginal evidence for the presence of two sub-populations in the BH mass spectrum with different redshift evolutions. Moreover, the two inferred redshift index posteriors overlap in the region consistent with the single redshift index of the common evolution model, hinting at the fact that the common evolution might simply be the average of the two evolving models.
\end{itemize}

The first point is crucial for population synthesis codes designed to describe the merging BBH population: comparing the outcome of such codes with hyperprior-driven results might result in a biased interpretation of the astrophysical processes at play. Whereas it is always true that while using parametrised models the possibility of introducing biases due to mismodelling is always present, in the specific case of the mass ratio distribution it is even more paramount: the lack of information carried by the data in a specific area of the parameter space makes it more difficult to discern between models yielding different behaviours.

A different redshift evolution is, to the best of our knowledge, the most robust smoking gun to reveal the presence of multiple sub-populations in the BH mass distribution: whereas other distinctive features may or may not be unique to specific formation channels, the fact that the redshift evolution can be different for different regions of the mass spectrum suggests that the timescales of the processes involved, and thus the processes themselves, are different. Our work, following up on the analysis presented by \citet{vanson:2022}, suggests that we might be already sensitive to the presence of two distinct formation channels: an investigation of their possible nature will be the subject of a future work.

\begin{acknowledgements}
The authors thank Thomas Callister and the anonymous referee for useful comments on the manuscript. The authors acknowledges financial support from the European Research Council for the ERC Consolidator grant DEMOBLACK, under contract no. 770017, and from the German Excellence Strategy via the Heidelberg Cluster of Excellence (EXC 2181 - 390900948) STRUCTURES. This work was funded by the Deut\-sche For\-schungs\-ge\-mein\-schaft (DFG, German Research Foundation) – project number 546677095. GD acknowledges financial support from the National Recovery and Resilience Plan (PNRR), Mission 4 Component 2 Investment 1.4 - National Center for HPC, Big Data and Quantum Computing - funded by the European Union - NextGenerationEU - CUP B83C22002830001. The authors acknowledge support by the state of Baden-W\"urttemberg through bwHPC and the German Research Foundation (DFG) through grants INST 35/1597-1 FUGG and INST 35/1503-1 FUGG.

This research has made use of data or software obtained from the Gravitational Wave Open Science Center (gwosc.org), a service of the LIGO Scientific Collaboration, the Virgo Collaboration, and KAGRA. This material is based upon work supported by NSF's LIGO Laboratory which is a major facility fully funded by the National Science Foundation, as well as the Science and Technology Facilities Council (STFC) of the United Kingdom, the Max-Planck-Society (MPS), and the State of Niedersachsen/Germany for support of the construction of Advanced LIGO and construction and operation of the GEO600 detector. Additional support for Advanced LIGO was provided by the Australian Research Council. Virgo is funded, through the European Gravitational Observatory (EGO), by the French Centre National de Recherche Scientifique (CNRS), the Italian Istituto Nazionale di Fisica Nucleare (INFN) and the Dutch Nikhef, with contributions by institutions from Belgium, Germany, Greece, Hungary, Ireland, Japan, Monaco, Poland, Portugal, Spain. KAGRA is supported by Ministry of Education, Culture, Sports, Science and Technology (MEXT), Japan Society for the Promotion of Science (JSPS) in Japan; National Research Foundation (NRF) and Ministry of Science and ICT (MSIT) in Korea; Academia Sinica (AS) and National Science and Technology Council (NSTC) in Taiwan.
\end{acknowledgements}
\bibliographystyle{aa}
\bibliography{bibliography.bib}
\end{document}